\documentclass[aps,pre,showpacs,showkeys,twocolumn,superscriptaddress,
floats,floatfix,preprintnumbers]{revtex4-1}
\usepackage{amssymb,amsmath}
\usepackage{bm,url,graphics,subfigure,epsfig,rotating,float}

 \usepackage{ulem}
\usepackage{mathrsfs} \usepackage{color,soul}

\newcommand{\ma}[1]{\ensuremath{\mathbb{#1}}}

\def \dtwo {d_{\rm 2}}
\def \Dtwo {D_{2}}

\def \zast {z^{\ast}}

\def \gv {{\rm g}}

\def \mP {\mathcal{P}}

\def \rhop {\rho_{\rm p}}

\def \uu  {{\bm u}}
\def \vv  {{\bm v}}
\def \ff  {{\bm f}}
\def \oo  {{\bm \omega}}
\def \ueta {u_{\rm \eta}}
\def \teta {\tau_{\rm \eta}}
\def  \xx  {{\bm x}}

\def \p {p}
\def \cs {c_{\rm s}}
\def \Lx {L_x}
\def \Ly {L_y}
\def \Lz {L_z}

\def  \RR  {{\bm R}}

\def  \VR  {V_{R}}
\def  \VRa  {|V_{R}|}
\def  \VRn  {V_{\rm n}}

\def \taup {\tau_{\rm p}}

\def \curl {{\bm \nabla} \times}

\def \delt {\partial_t}

\newcommand{\bra}[1]{\left\langle #1\right\rangle}

\def \Rlambda  {\mbox{Re}_\lambda}
\def \Ma  {\mbox{Ma}}

\def \St  {\mbox{St}}
\def \Fr  {\mbox{Fr}}
\def \Ro  {\mbox{Ro}}
\def \Sto  {\mbox{St}_{\rm 1}}
\def \Stt  {\mbox{St}_{\rm 2}}

\def \Teddy {T_{\rm eddy}}

\def \kf  {k_{\rm f}}

\def \urms  {u_{\rm rms}}

\def \Np  {N_{\rm p}}

\def \mp {m_{\rm p}}

\newcommand{\Eq}[1]{Eq.~(\ref{#1})}
\newcommand{\Fig}[1]{Fig.~(\ref{#1})}

\newcommand{\bfig}{\begin{figure}}
\newcommand{\efig}{\end{figure}}
\newcommand{\bc}{\begin{center}}
\newcommand{\ec}{\end{center}}
\newcommand{\bea}{\begin{eqnarray}}
\newcommand{\eea}{\end{eqnarray}}

\def \apj {ApJ}

\begin{document} 
\title{Statistics of relative velocity for particles
settling under gravity in a turbulent flow} 
\author{Akshay Bhatnagar}
\email{akshayphy@gmail.com} 
\affiliation{Nordita, KTH Royal Institute
of Technology and Stockholm University, Roslagstullsbacken 23, 10691
Stockholm, Sweden}

\preprint{NORDITA 2020-004}

\begin{abstract} 
We study the joint probability distributions of
separation, $R$, and radial component of the relative velocity, $\VR$, of 
particles settling under gravity in a
turbulent flow. 
We also obtain the moments of these distributions and
analyze their anisotropy using spherical harmonics. 
We find that the
qualitative nature of the joint distributions remains the same as no
gravity case. Distributions of $\VR$ for fixed values of $R$ show a
power-law dependence on $\VR$ for a range of $\VR$, exponent of the
power-law  depends on the gravity. 
Effects of gravity are also manifested in the following ways:
(a) moments of the distributions are anisotropic; degree
of anisotropy depends on particle's Stokes number, but does not depend
on $R$ for small values of $R$.  
(b) mean velocity of
collision between two particles is decreased for particles having
equal Stokes numbers but increased for particles having different
Stokes numbers. For the later,
collision velocity is set by the difference in their settling
velocities.
\end{abstract}

\maketitle
\section{Introduction} 
\label{sec:intro}

Small-sized heavy particles suspended in a turbulent flow are found in
many natural phenomena. Some of the examples are dust storms,
water droplets in clouds, and astrophysical dust in the protoplanetary disk around a young
star. Collisions of these particles play an important role in 
many processes.
For instance, in clouds, small water 
droplets collide with each other and may coalesce to form larger
droplets.  Understanding this process of collision and coalescence is
important to understand the process of initiation of rain from warm
clouds~\cite{sha03,Pruppacher2010microphysics,gra+wan13}.  
The motion of these small droplets is determined by the following two forces acting 
on them: (1) a hydrodynamic drag force applied by the surrounding fluid; 
(2) an external force such as gravity in the case of cloud droplets.  
If radius $a$ of the droplet is very small compared to the Kolmogorov dissipation scale 
$\eta$ of the flow and its density $\rho_p$ is much larger than the density $\rho$ of underlying
fluid, then the equation of motion of such a
droplet can be written as: 
\begin{eqnarray} 
\tfrac{{\rm d}}{{\rm d}t} \xx &= \vv \/ \label{eq:dxdt}\,,\quad
\tfrac{{\rm d}}{{\rm d}t}\vv = \frac{1}{\taup}\left[ \uu({\xx,t})
- \vv \right] +\gv \hat{\bm z}\,.  
\end{eqnarray} 
Here $\xx$ and $\vv$ denote the position and velocity of a particle, respectively. 
$\taup = (2\rhop/9\rho)a^2/\nu$ is the
characteristic response time of the particle, here $\nu$ is the kinematic 
viscosity of the fluid. $\gv$ is
the acceleration due to gravity and $\hat{z}$ is the unit vector along vertically 
downward direction.   
$\uu(\xx,t)$ is the flow velocity at the position of particle.  
This model assumes that fluid-inertia corrections are small and neglects the 
hydrodynamic interaction between the particles.
{This model \Eq{eq:dxdt} also neglects the history term present in
the Maxey-Riley equation~\cite{max+ril83}. Recent studies of
Refs.~\cite{dai+ant14,dai15} shows that history term plays an
important role in the dynamics of heavy particles and its effects 
depend on the density ratio $\rhop/\rho$. This ratio
is $1000$ for  water droplets in the air, in this case,
the effects of history term are found to be very small~\cite{dai15}.}
The inertia of the particle is measured in terms of Stokes number
$\St=\tau/\teta$, where  $\teta$ is the Kolmogorov time scale of the
turbulent flow.  To measure the relative importance of turbulence and
gravity, we define Froude number $\Fr = \frac{\ueta}{\tau_\eta \gv}$,
where $\ueta=\eta/\teta$, zero gravity corresponds to $\Fr = \infty$.  For cloud
droplets, $\St$ and $\Fr$ depend on the mean rate of energy dissipation
per unit volume $\varepsilon$. Value of $\varepsilon$ varies a lot in
clouds~\cite{vai+yau2000} and hence $\St$ and $\Fr$ can have different
values in different clouds. Typically for $10$ to $50$ micrometer-sized 
droplets $\St$ roughly lies between $0.1$ to $2.5$ and $\Fr$ may
vary from $0.05$ to $0.3$ ~\cite{aya+ros+wan+gra08,gra+wan13}.
{Note that, $\St$ and $\Fr$ appear by the nondimensionalization 
of \Eq{eq:dxdt} by using $\eta$, $\teta$, and
$\ueta$. Many
studies use another dimensionless number called Rouse number $\Ro =
\urms/(\taup \gv)$~\cite{ros+par+etal16,good+etal14}, where $\urms$ is
the root-mean-square speed of the flow. For a given value of 
Reynolds number, $\Ro$ can be expressed in terms of $\St$ and $\Fr$.
In the present work, we keep the Reynolds number of the flow fixed and
vary $\St$ and $\Fr$ independently. 
} 

It is known that small particles
cluster in turbulence due to their inertia.  Small scale clustering is
measured in terms of the probability $P(R)$ of finding two particles
within a separation $R$. It is found that $P(R)\sim R^{\dtwo}$ as $R \to 0$
~\cite{bec2007heavy}, $\dtwo$ is called the spatial correlation
dimension. If $\dtwo$ is smaller than the spatial dimension $d$, more
particles are found at small separation compared to uniformly
distributed particles.  
Inertial particles can detach from the 
flow streamlines and nearby particles can have high relative velocity, 
this is referred as sling effect~\cite{fal+pum07}. 
This can be understood in terms of
singularities in the  velocity field of the particles
~\cite{wilkinson2005caustics,wilkinson2006caustic}.  
These singularities are called caustics. 
Recently, it has been shown theoretically for smooth-random
flows~\cite{gustavsson2011distribution,gus+meh14} and in the direct
numerical simulations (DNSs) of the turbulent
flows~\cite{perrin2015relative,bhatnagar2018statistics} that the
probability distribution functions (PDFs) of the radial component of
relative velocity $\VR$ at small separation, have a power-law tail.
The exponent of this power-law is related to the phase space correlation
dimension $\Dtwo$.  It is also shown that there exists a parameter $\zast$
such that the joint PDFs of separation $R$ and $\VR$ are independent
of $\VR$ for $\VR \ll \zast R$ and are independent of $R$ for $\VR \gg
\zast R$~\cite{bhatnagar2018statistics}. {Here, $R$ and $\VR$ are 
nondimensionalized by using $\eta$ and $\ueta$, respectively.
Parameter $\zast$ sets a
velocity scale for a given separation $R$ and is often referred as
the matching scale~\cite{gus+meh14}.}
Refs.~\cite{gustavsson2011distribution,gus+meh14} gives the following theoretical
expression for the joint PDF $\mP(R,\VR)$, for $R < 1$, obtained for smooth random
flows in the white noise limit:
\begin{equation} 
\mP(R,\VRa) \sim R^{d-1} 
\begin{cases} R^{\Dtwo-d-1}
&\text{for $\VRa \le \zast R$}\\
\left(\frac{\VRa}{\zast}\right)^{\Dtwo-d-1} &\text{for $\zast R < \VRa < \zast$} \\
0 &\text{for $\VRa > \zast$}
\end{cases} 
\label{eq:jpdf}
\end{equation}
Form of the joint distribution $\mP(R,\VRa)$ agrees well with the theoretical prediction
\Eq{eq:jpdf} for particles suspended in a turbulent flow~\cite{bhatnagar2018statistics}.   
Mean collision velocities of particles in a polydisperse suspension
are studied in ~\cite{jam+ray18}. It is found that the collision
velocities between different sized particles can be much higher
compared to collision velocities between equal-sized particles. 
 
Many studies have focused on clustering and relative velocities of the
particles in turbulent flows without gravity. Relatively less is known
about the effect of gravity on clustering and
statistics of the relative velocities of the particles. 
For instance, it is not known, how gravity affects the form of the 
joint distribution $\mP(R,\VRa)$ given in \Eq{eq:jpdf}? 
How anisotropy of the system changes as a function of $R$ and $\St$?
How mean collision velocities of the particles are
modified by the gravity? This paper focuses on these questions. 

Gravity and turbulence together can give rise to many non-trivial phenomena
in the dynamics of the heavy particles. 
{In the presence of turbulence, 
particles are found to settle with a speed that is higher compared to their 
terminal-speed in still fluid~\cite{max87,wan+max93,bec+hom+ray14}. 
Reduction of settling speed by turbulence is also observed in the
experiments~\cite{good+etal14}. Simulations with non-linear drag force
on the particles show a similar reduction in the settling speeds.
Present work considers only the linear drag case 
(for which mostly enhancement is observed in simulations), and does 
not address the issue of enhancement vs hindering of settling by 
turbulence.}

Refs.~\cite{gus+vaj+meh14,bec+hom+ray14,ire+bra+col16} shows
that the presence of gravity modifies the small scale
clustering of heavy particles. {This modification depends on the values
of $\St$ and $\Fr$.
For particles having $\St < 1$ clustering is reduced compared to 
no-gravity case whereas for particles having $\St>1$ clustering is
significantly increased compared to no-gravity case. 
In a real experiment, gravity is always
present hence,
these results can be used to understand the experimental observation
of particle clustering in turbulent flows~\cite{sumbekova_PRF17,petersen_JFM19,obligado_JOT14}. } 
It is argued that
modification of clustering happens because settling particles sample
the flow differently compared to particles advected solely by
turbulence.  

Mean relative velocity of the particles having equal $\St$ is found 
to be reduced by gravity~\cite{bec+hom+ray14,ire+bra+col16}. 
Refs~\cite{pari+aya+ros+wan+gra15,ire+bra+col16} study the PDFs of 
relative velocity at small $R$ in the presence of gravity.  
These studies find that the PDFs remains non-Gaussian similar to the
case when $\Fr=\infty$ but, fluctuations are reduced due to the presence of the gravity. 
Mean collision rates
of equal $\St$ particles are found to be reduced by 
gravity~\cite{aya+ros+wan+gra08,oni+tak+kom09,woi+jon+por09}. This is
due to the significant reduction of the mean relative velocity.  
Anisotropy of the clustering and mean relative velocity is analyzed in
Ref.~\cite{ire+bra+col16} by calculation the radial distribution
function and mean relative velocity as a function of $R$ and angle 
$\theta$ between separation vector and direction of gravity. Using
spherical harmonic decomposition of these quantities it is shown that
the anisotropy is significant at small separation for large values of
$\St$. This analysis of anisotropy is done for separations larger than $\eta$. 
Statistics of relative velocities in bi-disperse turbulent
suspensions are studied in Ref~\cite{dha+bra18}. It is found that
gravity enhances the relative velocities in vertical and horizontal
directions. It is shown that for small values of $\Fr$ relative velocity
is dominated by differential settling but turbulence still plays an
important role.    

As radii of droplets are always smaller than the dissipation scale
$\eta$, the separation between the droplets is also much smaller than
$\eta$ when they collide.  Therefore, we study the relative velocities of particles for
small values of separations.  In this paper, we focus on the joint PDFs
$\mP(R,\VRa)$ and its moments for different values of $\St$ and $\Fr$.
We use length scale $\eta$ and velocity scale $\ueta$ to nondimensionalize separation $R$ 
and  relative velocity $\VR$, respectively.
For separation $R<1$.  We show the following results for the first
time: 
\begin{itemize} 
\item The joint PDF $\mP(R,\VRa)$, is qualitatively similar to the case of zero
gravity; in the sense that in both of these cases: 
\begin{itemize}
\item  There exists a parameter $\zast$ such that at small $R$ the joint
PDF is independent of $\VR$ for $\VR < \zast R$,  
and is independent of $R$ for $\VR >  \zast R$.  
\item For a fixed $R$ the PDF as a function
of $\VR$ shows a power-law range with exponent $\Dtwo - 4$.
Although the phase-space correlation dimension, $\Dtwo$, itself is not
the same as the zero-gravity case.  
\end{itemize} 
\item The spatial
clustering, described by the $0$-th moment of the joint PDF, is
anisotropic, in a very special way. A decomposition of the moment into
spherical harmonics show that different order harmonics, for $\ell =
0,2,$ and $4$, scales with the same exponent $\dtwo$ but the
amplitudes depend on $\ell$.   
\item Similar behaviour is seen for
the first moment, which describes the mean relative velocities, too.
\item We define collision velocities as the relative velocity of two
particles separated by the sum of their radii. The mean and RMS of
collision velocities as a function of $\Sto$ and $\Stt$ changes
qualitatively from the zero-gravity case; in particular the contours
in the $\Sto-\Stt$ plane becomes parallel to the diagonal as $\Fr$
decreases, i.e., the collision velocities are a function of
$\mid\Sto-\Stt\mid$ alone.  
\item Furthermore, for particles with
equal $\St$, the mean collision velocity decreases as $\Fr$ decreases,
i.e., gravity increases. This decrease is more pronounced at higher
$\St$ than at lower ones.  
\item For $\Sto \neq \Stt$ the
qualitative behaviour is opposite, the mean collision velocity
increases as $\Fr$ decreases.  In this case, mean collision velocity is
set by the difference in the settling velocities.
\end{itemize}

\section{Direct numerical simulation} 
\label{sec:dns}

The flow velocity $\uu(\xx,t)$ is determined by solving the
Navier--Stokes equation 
\begin{subequations} 
\begin{align}
\tfrac{\partial }{\partial t}\rho &+ \nabla \cdot (\rho \uu) = 0
\/\,, \label{eq:density}\\ 
\rho \tfrac{{\rm D} }{{\rm D}t}\uu &=
-\nabla \p + \mu \nabla \cdot {\ma S} + {\bm f}  \,.  \label{eq:mom}
\end{align} 
\label{eq:fluid} 
\end{subequations} 
Here $\tfrac{\rm
D}{{\rm D}t} \equiv \delt + \uu \cdot \nabla$ is the Lagrangian
derivative,  $\p$ is the pressure of the fluid, and $\rho$ is its
density as mentioned above.  The dynamic viscosity is denoted by
$\mu\equiv\rho\nu$, and ${\ma S}$ is the second-rank tensor with
components $S_{kj} \equiv \partial_k u_j + \partial_j u_k
-\delta_{jk}(2/3)\partial_k u_k$ (Einstein summation convention).
Here $\partial_k u_j $ are the elements of the matrix $\ma A$ of
fluid-velocity gradients. $\ff$ denotes the external force.
To relate pressure $\p$ and density $\rho$,  
we use the ideal gas equation of state with a constant speed of sound, $\cs$.

To solve Eqs.~(\ref{eq:fluid}) we use an open-source code called
The Pencil Code~\cite{pencil-code}. This code has been used earlier
for the similar studies of particles-laden turbulent flows in
Refs~\cite{bhatnagar2018statistics} and ~\cite{bhatnagar2018bid}. 
It uses a
sixth-order finite-difference scheme to compute space derivatives and a third-order 
Williamson-Runge-Kutta~\cite{wil80} scheme for time evolution.
We use periodic boundary conditions in all three directions.
A white-in-time, Gaussian, forcing $\ff$ is used
that is concentrated on a shell of wavenumber with radius
$\kf$ in Fourier space~\cite{B01}. Forcing term is integrated by using the
Euler--Marayuma scheme~\citep{hig01}.  
A statistically stationary state is reached, where 
the average energy injection by the external forcing $\ff$
is balanced by the average
energy dissipation by viscous forces.  
The amplitude of the
external forcing is chosen such that the Mach number, $\Ma \equiv
\urms/\cs$ is always less than $0.1$, i.e., the flow is weakly
compressible, where $\urms = \sqrt{\bra{\uu^2}}$ is the
root-mean-squared velocity of the flow. This week compressibility has no 
important effect on our results; please see
the discussion in Ref.~\cite{bhatnagar2018statistics}, section II and
Appendix A in Ref~\cite{bhatnagar2018statistics} for further details.
The same code with similar setting has been used before to study the scaling and
intermittency in fluid and magnetohydrodynamic
turbulence~\cite{dob+hau+you+bra03,hau+bra+dob03,hau+bra04}.
Our simulations are
performed in a three-dimensional periodic box with sides
$\Lx=\Ly=2\pi$ and $\Lz=4\pi$. This box is discretised in $N$ equally
spaced grid points in each direction. 

Due to gravity and periodic boundary conditions, settling particles
can travel box length $\Lz$ in a time interval which is less compared to the
correlation time scale of the flow, $\Teddy$ in our case. 
This can
produce errors in the statistics measured for the
particles~\cite{pari+aya+ros+wan+gra15,ire+bra+col16}. To minimize this,
we use a longer box in the $z$ direction which is the direction of gravity.  
From the values of the parameters in TABLE~\ref{table:para}, we
estimate that having $\Lz=4\pi$ is good enough for the values of $\St$
considered in this paper.

We introduce the particles into the simulation after the flow has reached
a statistically stationary state.  
Initially,  particles are uniformly distributed in the simulation domain
with zero initial velocities.  
To evolve positions and velocities of particles according to \Eq{eq:dxdt}, 
we use a third-order Runge-Kutta scheme.  
To obtain the flow velocity at the positions of the particles,
we use a tri-linear interpolation method.

\begin{table} 
\begin{center} 
\caption{Parameters for our DNS runs with
$N^3$ collocation points,  $\Np$ is the number of particles.  Further,
$\nu$ is the kinematic viscosity, $\epsilon$ in the mean rate of
energy dissipation, $\eta \equiv (\nu^3/\epsilon)^{1/4}$ and $\teta =
(\nu/\epsilon)^{1/4}$ are the Kolmogorov length and time scales. These
numbers are quoted in dimensionless units (see text).  The Reynolds
number $\Rlambda$ is based on the Taylor microscale $\lambda$, and $\Teddy$ is 
large eddy turn-over time scale of the flow (see text).}
\label{table:para} 

\begin{tabular}{c c c c c c c c} 
\hline\hline 
$N$ & $\nu$ & $\Np$ & $\Rlambda$ & $\varepsilon$ & $\eta$ & $\teta$ & $\Teddy$ \\ 
\hline 
$512$ & $5.0\times 10^{-4}$ & $10^7$ & $90$ & $3.25\times 10^{-3}$ &
$1.4\times 10^{-2}$ & $0.39$ & $0.86$ \\ 
\hline 
\end{tabular}
\end{center} 
\end{table}
Parameters of the simulations are given in TABLE~\ref{table:para}.
$\varepsilon \equiv 2\nu \Omega$ is 
the mean rate of energy dissipation per unit volume, where
the enstrophy $\Omega \equiv \bra{|\oo|^2}$, and $\oo \equiv \curl
\uu$ is the vorticity.
Taylor microscale Reynolds number is defined as $\Rlambda\equiv
(\urms\lambda)/\nu$, where
$\urms$ is the root-mean-square velocity of the flow averaged over the
whole domain, $\lambda\equiv\sqrt{(5\urms^2)/(2\Omega)}$ is the Taylor
microscale of the flow, and $\nu$ is the kinematic viscosity.  
  The Kolmogorov length  is defined as $\eta
\equiv (\nu^3/\varepsilon)^{1/4}$, the characteristic time scale of
dissipation is given by $\teta = (\nu/\varepsilon)^{1/4}$ and $\ueta
\equiv \eta/\teta$ is the characteristic velocity scale at the
dissipation length scale.  
The large eddy turnover time is given by $\Teddy\equiv 1/(\kf\urms)$.
In rest of this paper, unless otherwise stated,
we use $\eta$, $\teta$, and $\ueta$ to non-dimensionalize length,
time, and velocity respectively.  

\section{Results} 
\label{sec:res}

We compute the joint PDFs $\mP(R,\VRa)$ and its $p-$th moments defined as:
\begin{eqnarray} \mp(R) = \int \VRa^{p} \mP(R,\VRa) {\rm d}\VRa.
\end{eqnarray} 
Here, $R$ and $\VR$ are nondimensionalized by scales $\eta$ and $\ueta$.  
$m_0(R)$ measures the number of particle pairs having
separation between $R$ and $R+{\rm d}R$. If there is no clustering
$m_0(R)\sim R^2$ in three dimensions. 
We define $\VRn$ as $\VRa$ when two particles are moving towards each
other ($\VR < 0$) and separation  $R$ between their centre of masses is equal to 
$a_1+a_2$, where $a_1$ and $a_2$ are the radii of two particles
nondimensionalized by $\eta$. Please note that the actual parameters
that characterize the particles are $\Sto$ and $\Stt$. We fix the
density ratio $\rhop/\rho=10^3$ (which is the case for the water droplets
in the air) to get the radii $a_1$ and $a_2$ from
$\Sto$ and $\Stt$, respectively
\footnote{Radius of the particles is given by $a/\eta =
\sqrt{\frac{9}{2}\frac{\St}{\rhop/\rho}}$}. 
We call $\VRn$
the collision velocity and consider it as a proxy for actual collision velocity.
Mean of $\VRn$ is defined as:
\begin{eqnarray}
\bra{\VRn} =  \frac{\int_{-\infty}^0 \VRa \mP(R,\VRa) {\rm d}\VR}{\int_{-\infty}^0
\mP(R,\VRa){\rm d}\VR} \mid_{R=a_1+a_2}.
\label{eq:mVn}
\end{eqnarray}

\subsection{Real space clustering} \label{subsec:clus}

\begin{figure} 
\begin{center}
\includegraphics[width=1.0\linewidth]{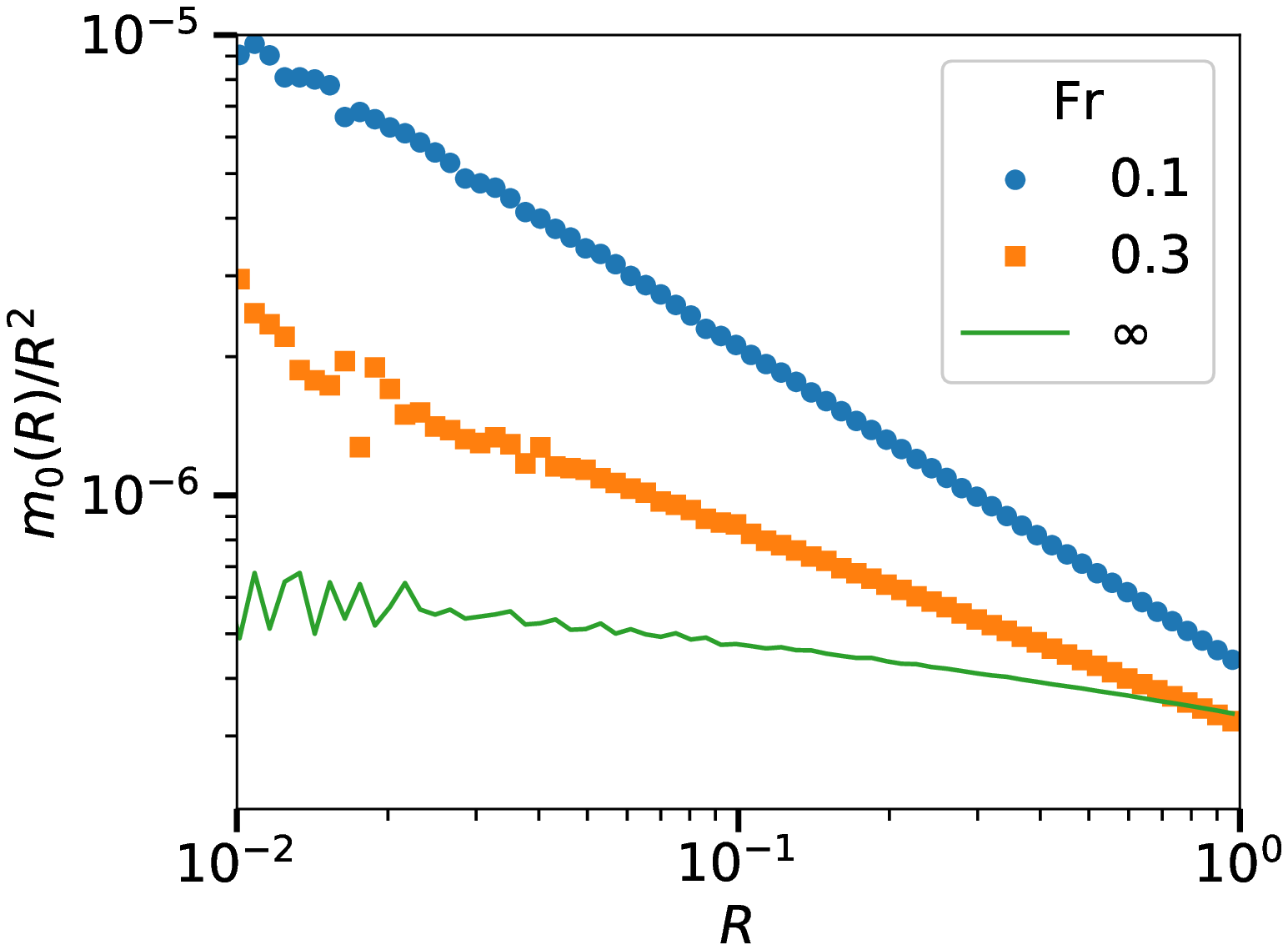}
\put(-32,165) {\colorbox{white}{(a)}}\\
\includegraphics[width=1.0\linewidth]{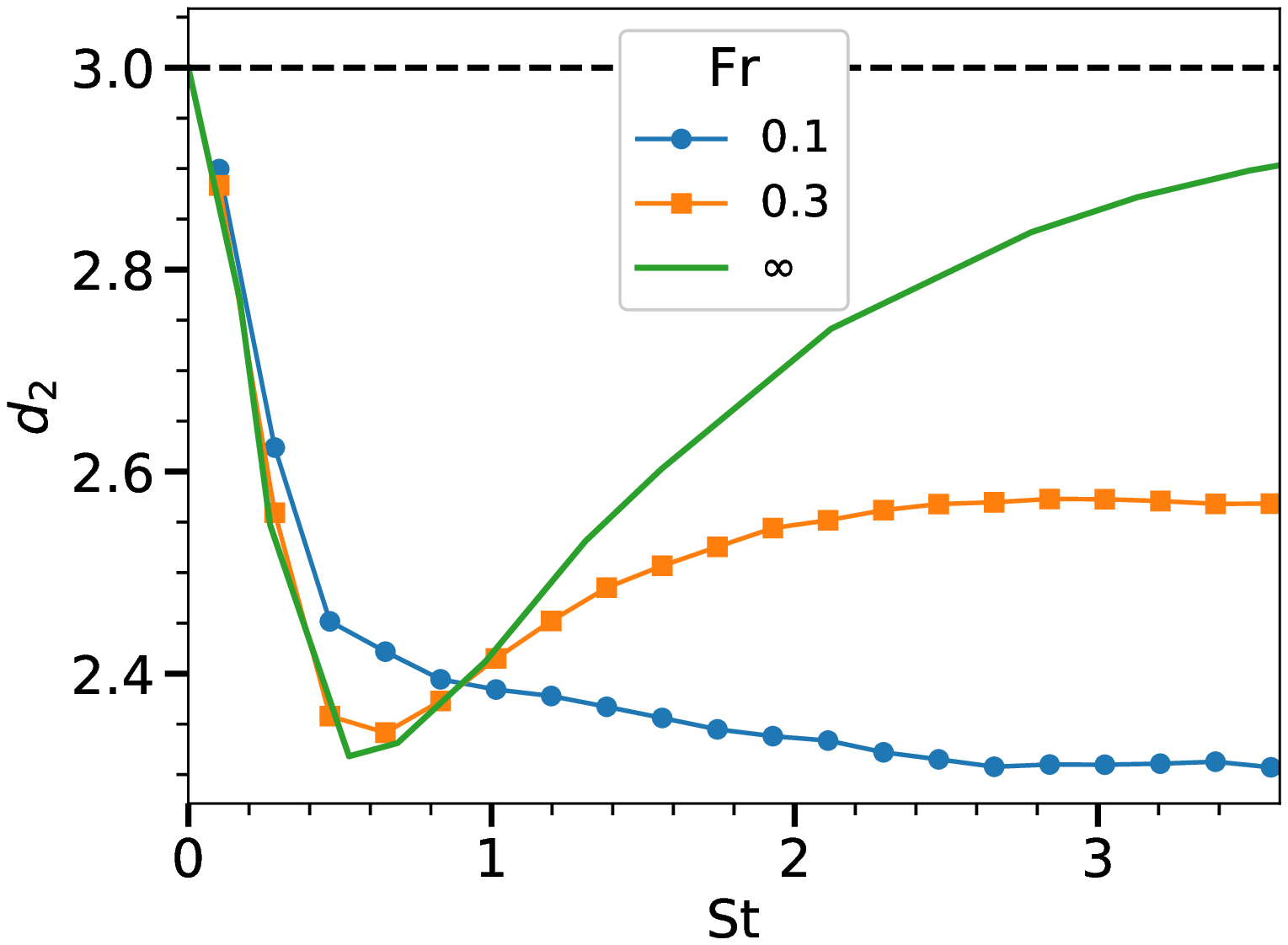}
\put(-30,165) {\colorbox{white}{(b)}} 
\caption{(color online) (a)
$m_0(R)/R^2$ as a function of $R$ for $\St=3.02$ and three different
values of $\Fr$.  (b) Real space correlation dimension $\dtwo$ as a
function of $\St$ for three different values of $\Fr$.}
\label{fig:d2Fr} 
\end{center} 
\end{figure}

\Fig{fig:d2Fr}(a) shows plots of $m_0(R)/R^2$ as a function of $R$ for
$\St=3.02$ and three different values of $\Fr$. We observe that
$m_0(R)$ has a power-law dependence on $R$ with exponent equals to
$\dtwo-1$.  For $\Fr=\infty$ data shows no clustering as $m_0(R)/R^2$
is independent of $R$. We observe that for the same $\St$ but $\Fr=0.3$, and
$0.1$ there is a significant amount of clustering.  In
\Fig{fig:d2Fr}(b), we plot real space correlation dimension $\dtwo$ as
a function of $\St$ for three different values of $\Fr$.  We find that
for $\Fr=0.1$ and $0.3$ and $\St>1$, $\dtwo$ is small compared to
$\Fr=\infty$. This implies that the settling enhances the clustering
for $\St>1$ for the values of $\Fr$ considered here. This is
consistent with the results of Refs.~\cite{bec+hom+ray14}. 
{We also observe that for $\Fr=0.1$ and $0.3$, $\dtwo$ becomes constant
for large values of $\St$. This indicates that the clustering does not
depends on $\St$ for large values of $\St$, this is
consistent with the experimental observation of
Refs~\cite{sumbekova_PRF17,petersen_JFM19}.} For the
range of $\St$ and $\Fr$ studied here, $\dtwo$ remains less than $d$, 
hence the phase space correlation dimension $\Dtwo=\dtwo$. 

\subsection{Joint distribution of $R$ and $\VR$} 
\label{subsec:jpdfs} 
 
\begin{figure} 
\begin{center}
\includegraphics[width=1.0\linewidth]{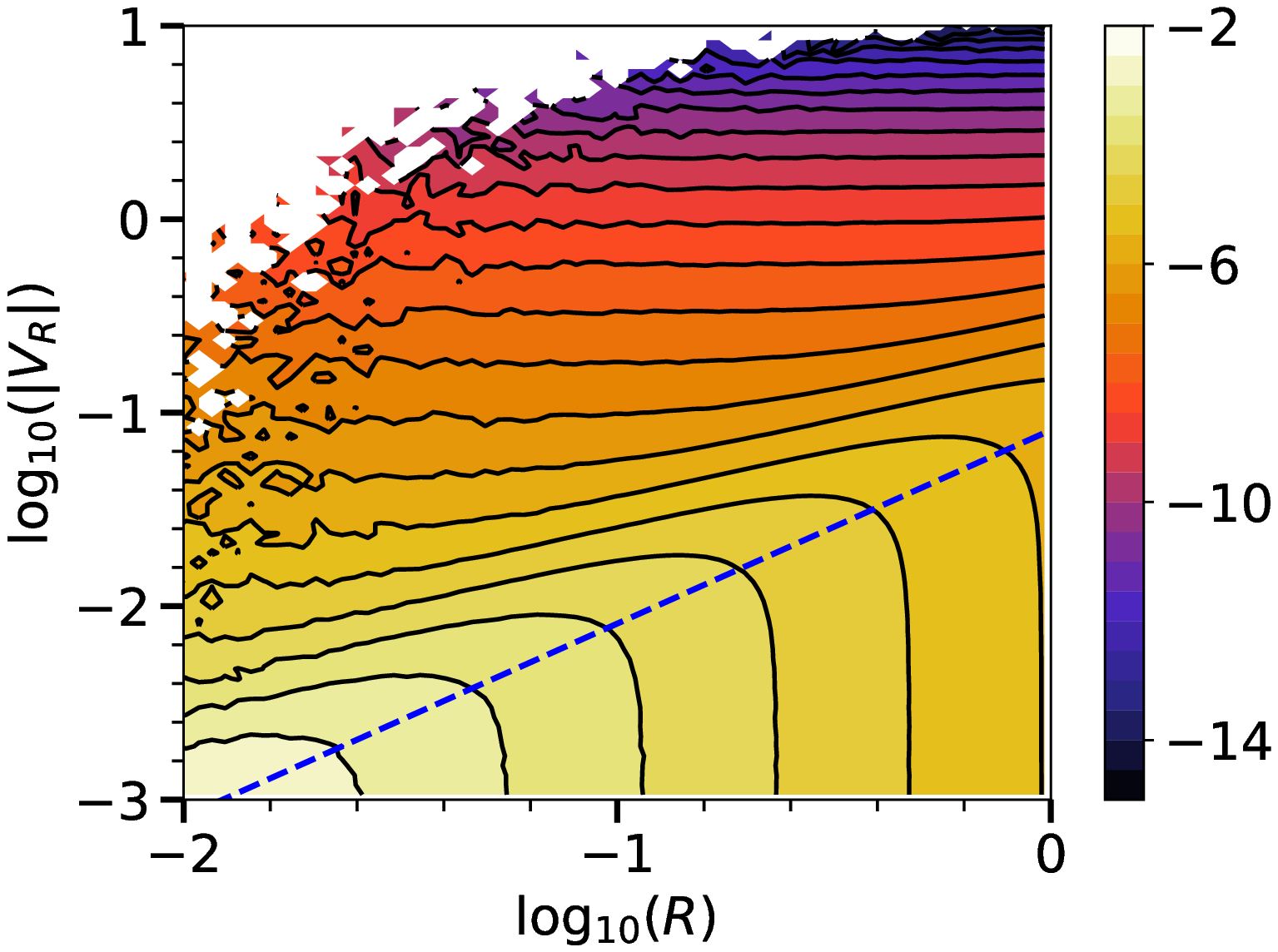}
\put(-62,165) {\colorbox{white}{(a)}}\\
\includegraphics[width=1.0\linewidth]{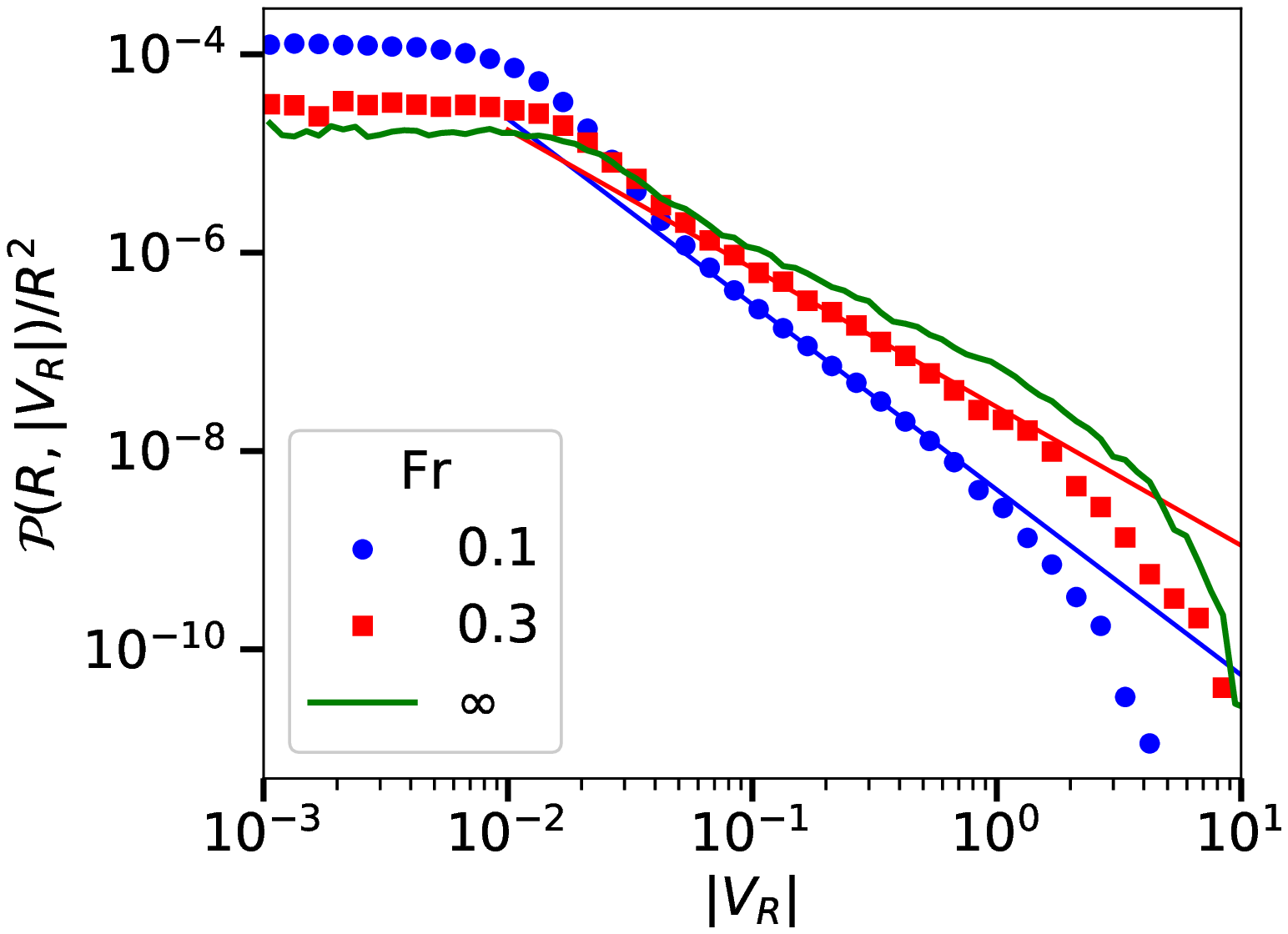}
\put(-35,165) {\colorbox{white}{(b)}} 
\caption{(color online) 
(a) Contour plots of $\log_{10}{(\mP(R,\VR)/R^2)}$ for $\St = 2.11$ and
$\Fr = 0.1$.  (b) $\mP(R,\VR)/R^2$ for $R=0.1$, $\St=2.11$, plotted
for three different values of $\Fr$. Solid lines in (b) have slopes equal to $\Dtwo-4$.} 
\label{fig:jpdfRo1}
\end{center} 
\end{figure}

\Fig{fig:jpdfRo1} (a) shows the joint distribution $\mP(R,\VRa)/R^2$
for $\St=2.11$ and $\Fr=0.1$. 
The dashed blue line shows $\VRa=\zast R$, where $\zast\simeq 0.1$ is found by fitting
a line to the data. We find that the value of $\zast$ does not depends
much on $\St$, this is consistent with the results of ~\cite{bhatnagar2018bid}.
We also observe that $\zast$ has a very week dependence on $\Fr$ but,
we do not have many values of $\Fr$ to comment on this dependence.  
We observe that for $\VRa < \zast R$,
contour lines are vertical which implies that the distribution depends 
only on the separation $R$. 
For $\VRa > \zast R$ contour lines are horizontal, 
implying that it depends only on $\VRa$ in this regime.
Similar nature of the distributions $\mP(R,\VRa)$ is observed for  
zero gravity ($\Fr = \infty$) case in Refs.~\cite{bhatnagar2018statistics,bhatnagar2018bid}.  
In \Fig{fig:jpdfRo1} (b), we plot the
distribution $\mP(R,\VR)$ for $R=0.1$, $\St=2.4$, and for three
different values of $\Fr$. We find that same as no-gravity case
distributions have a power-law tail with scaling exponent $\Dtwo-4$.
As $\Dtwo$ changes with $\Fr$ see \Fig{fig:d2Fr}, values of scaling
exponents also changes. Observe that as we decrease $\Fr$, power-law
tails of the distributions become steeper. This implies that the mean
and rms values of $\VRa$ also decrease as gravity increases.   

{To understand collisions of particles one would like to
know the distribution of $\VR$ at $R=2a$, where $a$ is the radius of the 
particle. For a given value of $\St$, the radius depends on the density ratio $\rhop/\rho$ 
(as described at the beginning of Section~\ref{sec:res}). For example,
for $\St=2.11$ and $\rhop/\rho=10^3$, value of $a/\eta$ is roughly
$0.1$, hence data for $R<0.1$ in \Fig{fig:jpdfRo1} (a) is irrelevant
for this density ratio. For a different value of $\rhop/\rho$ and the same
value of $\St$, one would need the distribution at a different value of
$R$, therefor covering a range of $R$ is useful to know the statistics
of $\VR (R=2a)$ for a range of $\rhop/\rho$.}

\subsection{Effect of anisotropy} 
\label{subsec:aniso} 
\begin{figure*}
\begin{center}
\includegraphics[width=0.32\linewidth]{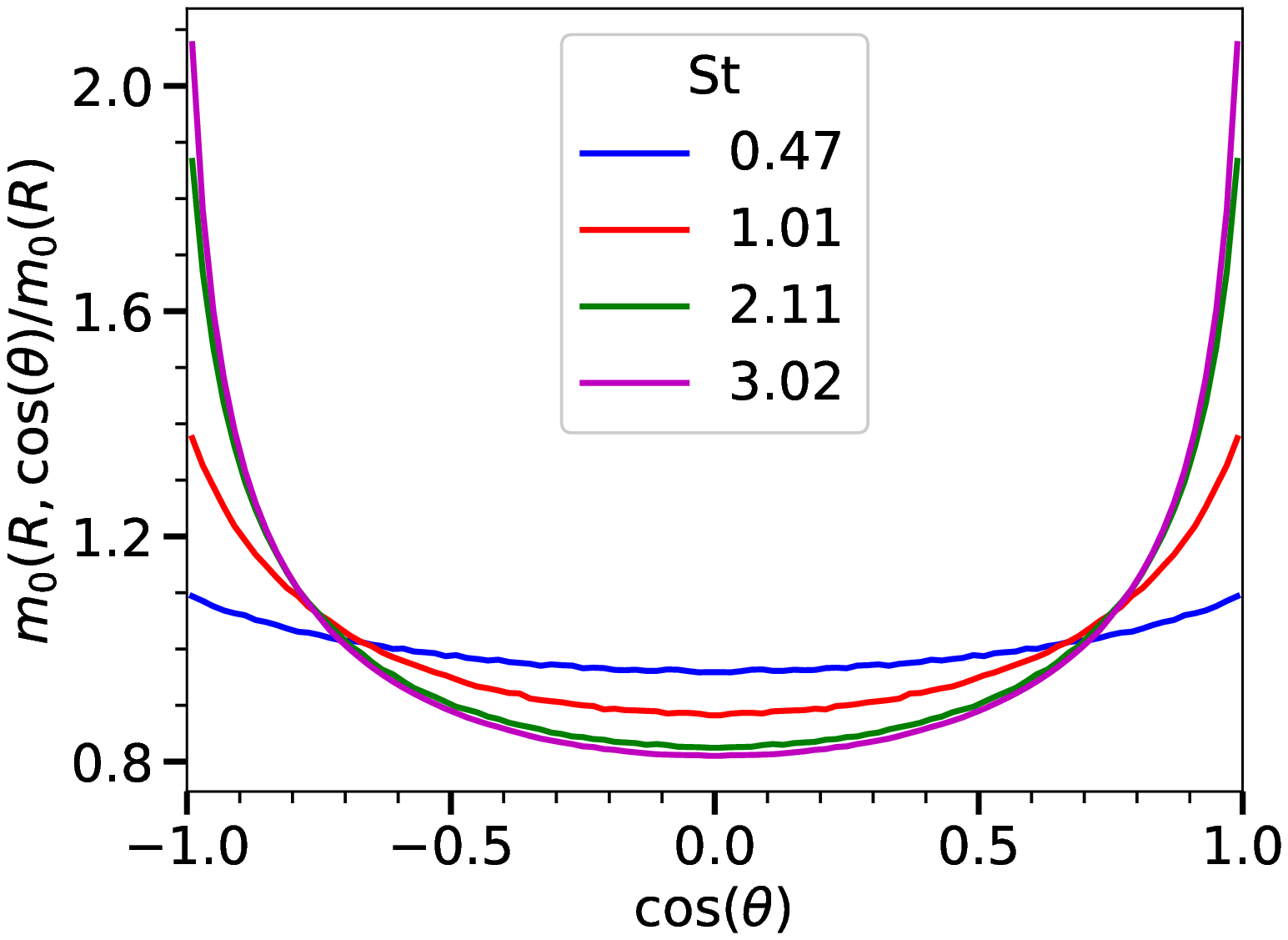}
\put(-30,90) {\colorbox{white}{(a)}}
\includegraphics[width=0.32\linewidth]{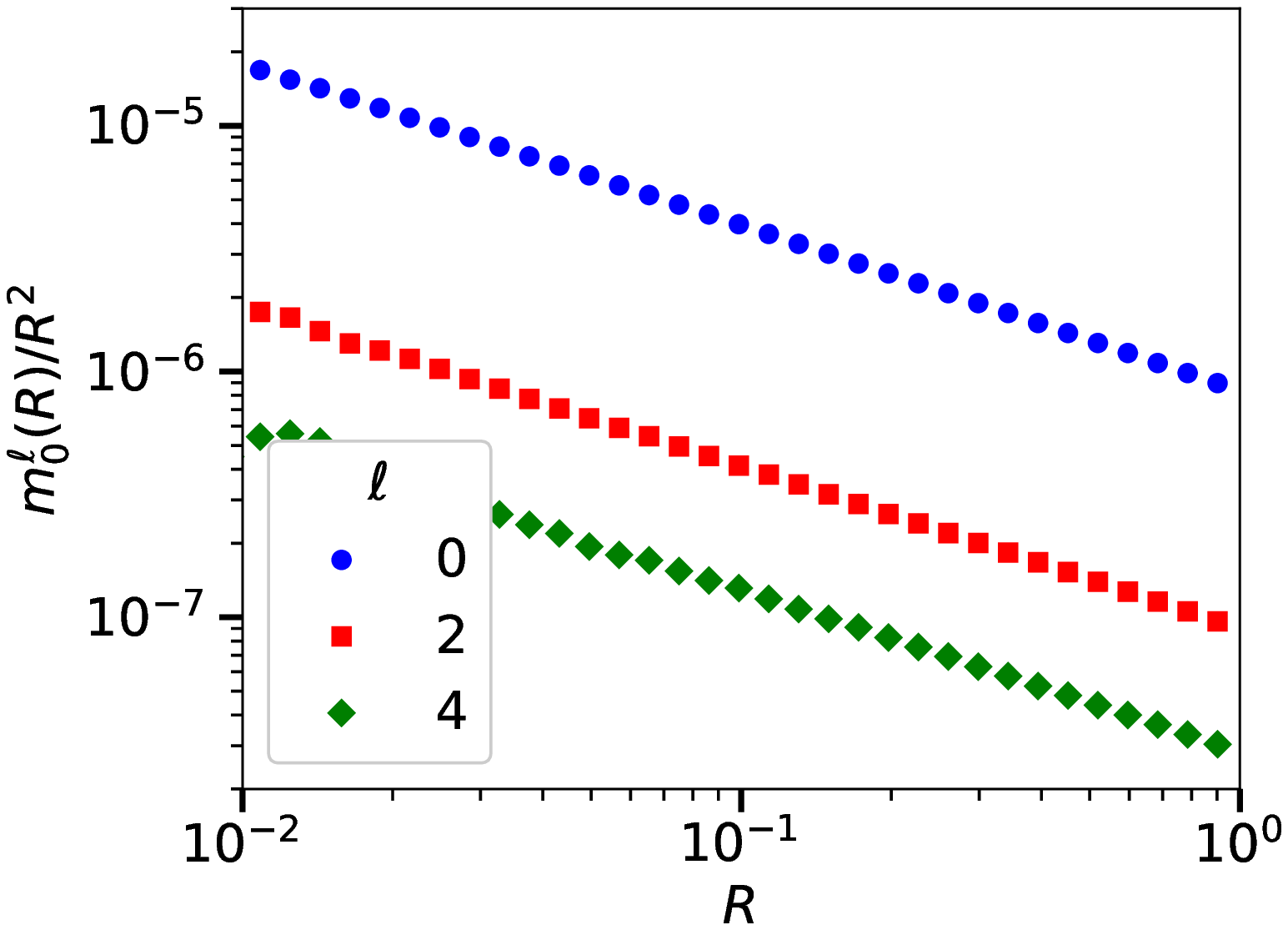}
\put(-30,90) {\colorbox{white}{(b)}}
\includegraphics[width=0.32\linewidth]{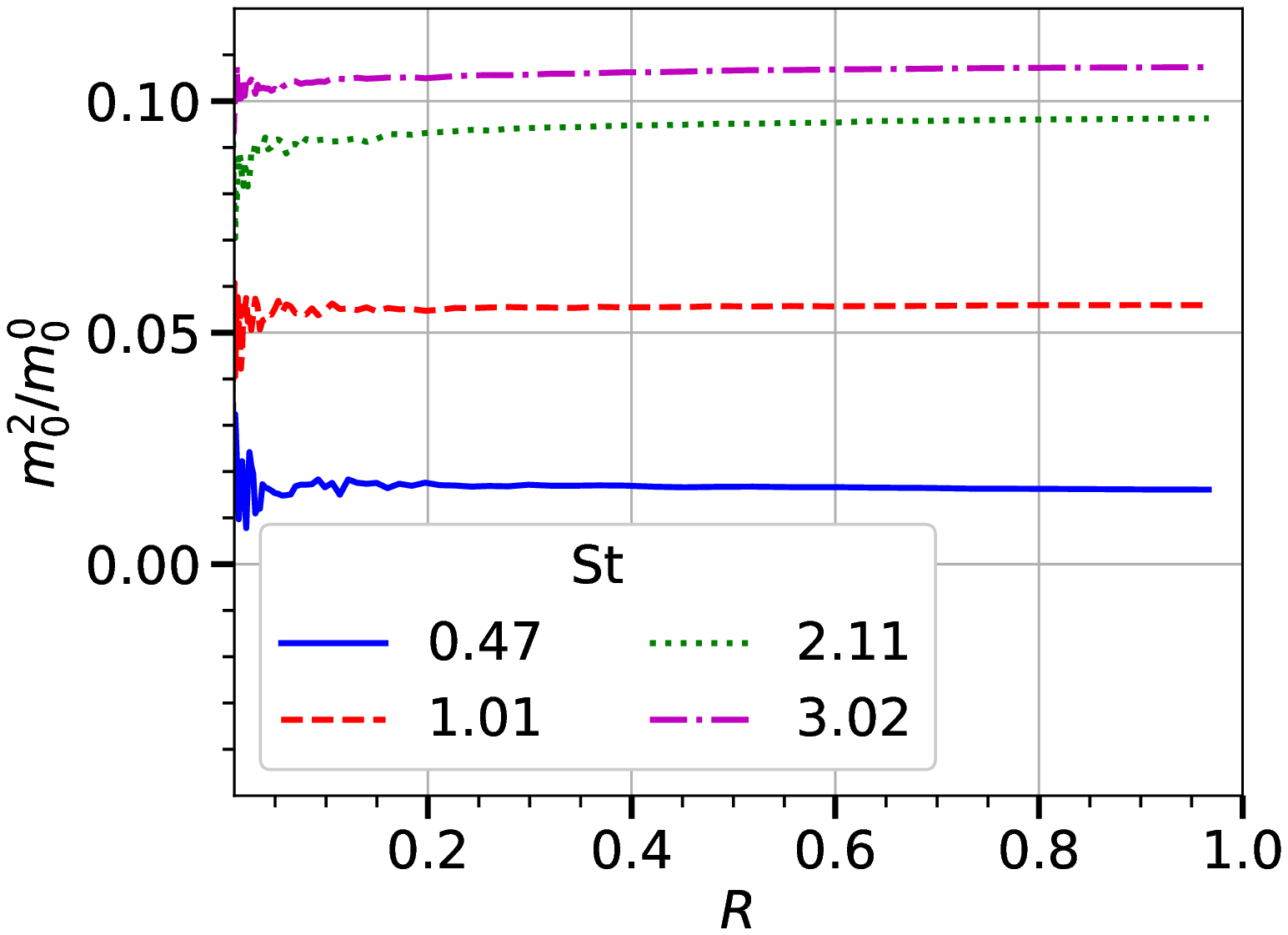}
\put(-30,90) {\colorbox{white}{(c)}} 
\caption{(color online) (a)
$0$-th moment as a function of $\cos{(\theta)}$ for $R=0.5$ normalized
by its average over $\cos{(\theta)}$ for $\Fr=0.1$ and four different
values of $\St$.  (b) $m^\ell_0$ for $\ell=0,2$ and $4$ plotted as a
function of $R$ for $\St=3.02$.  (c) measure of anisotropy
$m^2_0/m^0_0$ plotted as function of $R$ for $\Fr=0.1$ and four
different values of $\St$.  } \label{fig:m0Rcosth} \end{center}
\end{figure*}
\begin{figure*} 
\begin{center}
\includegraphics[width=0.32\linewidth]{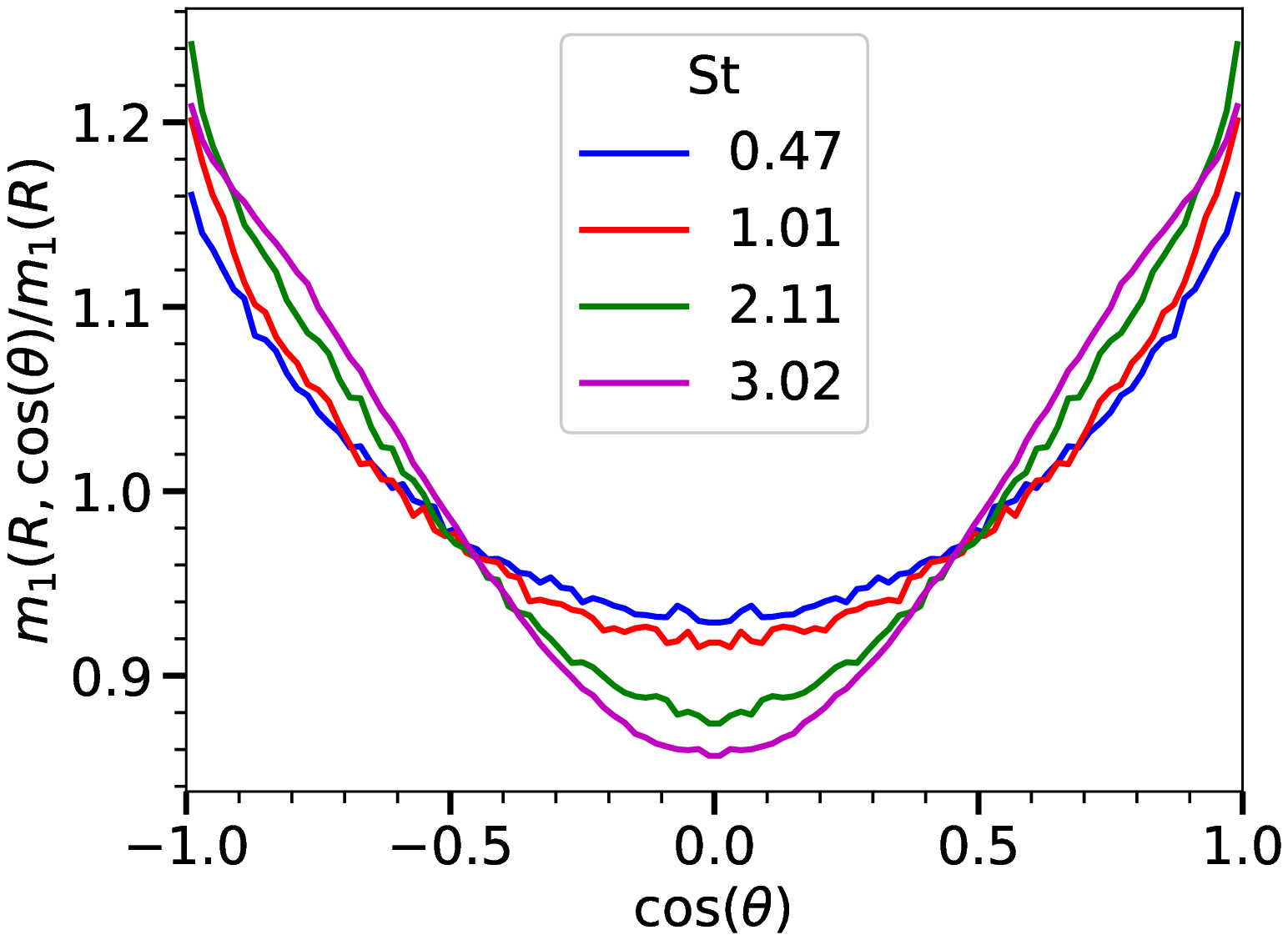}
\put(-130,105) {\colorbox{white}{(a)}}
\includegraphics[width=0.32\linewidth]{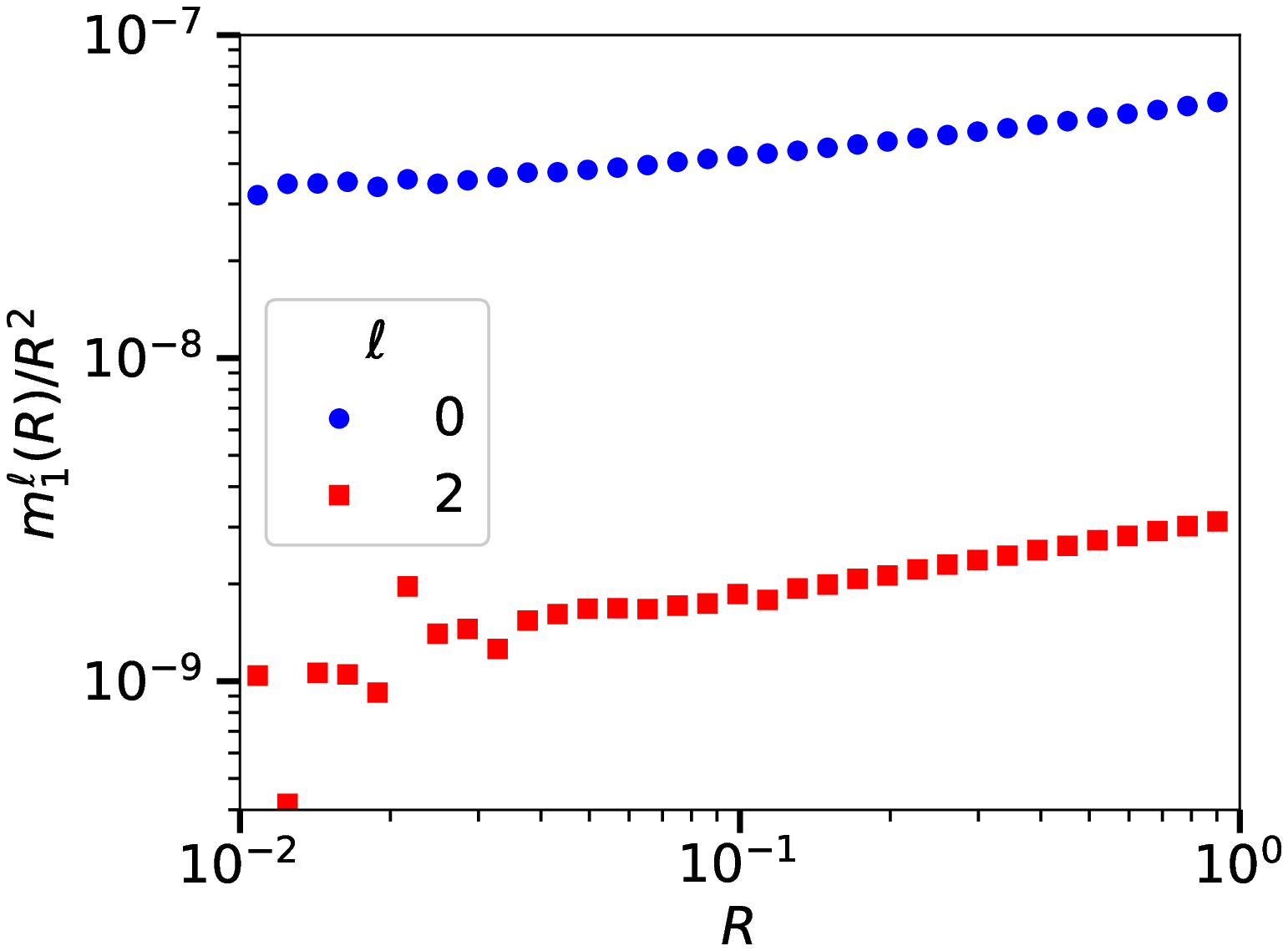}
\put(-130,105) {\colorbox{white}{(b)}}
\includegraphics[width=0.32\linewidth]{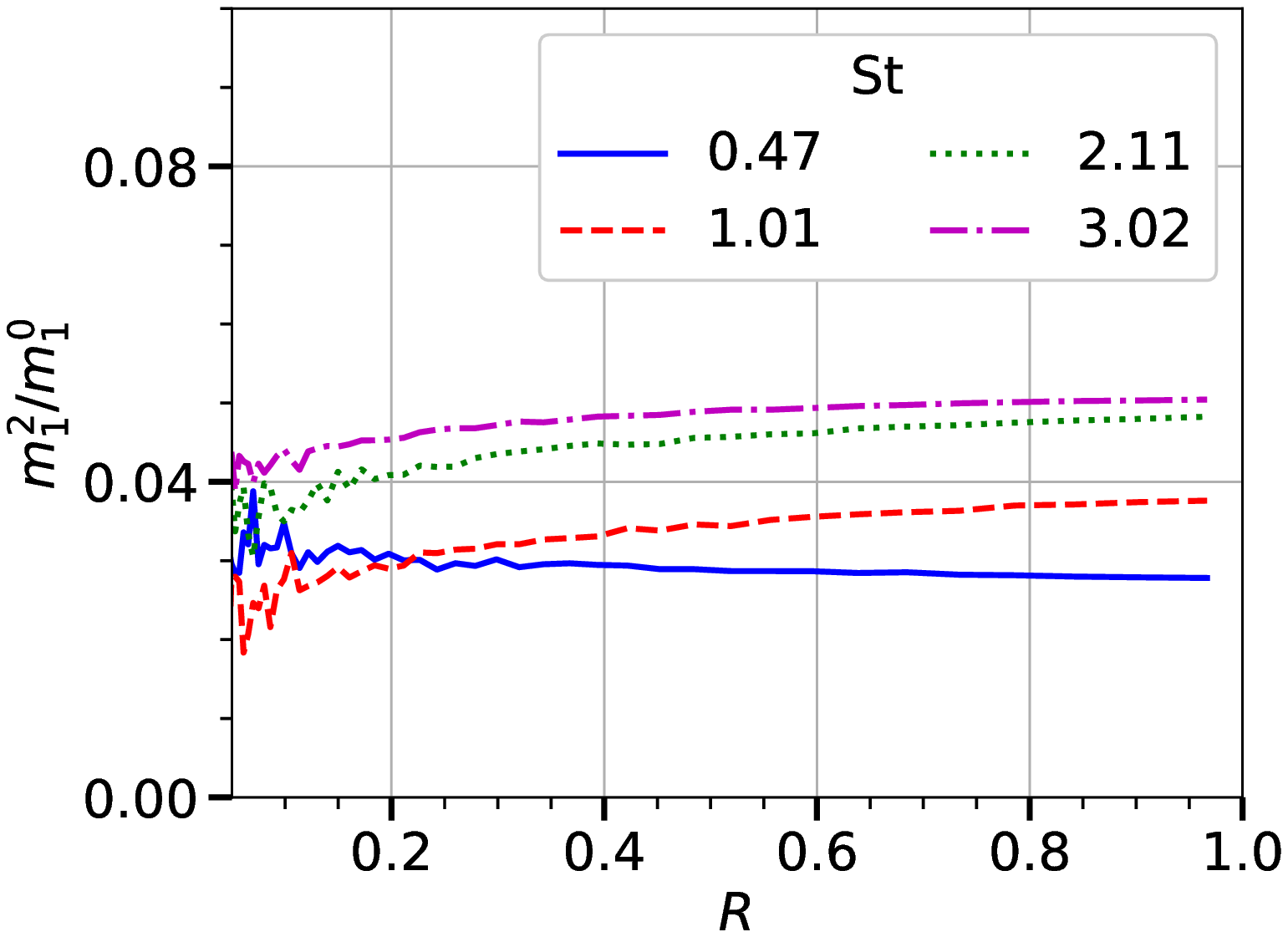}
\put(-130,105) {\colorbox{white}{(c)}} 
\caption{(color online) (a)
$1$-th moment as a function of $\cos{(\theta)}$ for $R=0.5$ normalized
by its average over $\cos{(\theta)}$ for $\Fr=0.1$ and four different
values of $\St$.  (b) $m^\ell_1$ for $\ell=0$ and $2$ plotted as a
function of $R$ for $\St=3.02$.  (c) measure of anisotropy
$m^2_1/m^0_1$ plotted as function of $R$ for $\Fr=0.1$ and four
different values of $\St$.  } \label{fig:m1Rcosth} \end{center}
\end{figure*}
Above calculations do not take in to account the anisotropy due to
the presence of gravity.  To compute the anisotropic contributions, we can
obtain the joint distribution $\mP(R,\VRa,\cos(\theta))$, where
$\theta$ is the angle between $\RR$ and direction of gravity
$\hat{z}$. We compute the moments $\mp(R,\cos{(\theta)})$ of this
distribution and use spherical harmonics decomposition.
{A function $f(R,\theta,\phi)$ that depends on spherical co-ordinates
$R$, $\theta$, and $\phi$ can be expressed as a linear
combination of Spherical harmonics $Y_\ell^m(\cos(\theta),\phi)$ as:}
\begin{eqnarray}
 f(R,\theta,\phi) = \sum_{\ell=0}^\infty \sum_{m=-\ell}^\ell
f_\ell^m(R) Y_\ell^m(\cos(\theta),\phi)
\label{eq:Ylm}
\end{eqnarray} 
As there is no dependence on the azimuthal angle $\phi$ in our system, only $m=0$
modes are present.  We can write: 
\begin{eqnarray} 
\mp(R,\cos(\theta))
= \sum_{\ell}\mp^\ell(R) P_\ell(\cos{(\theta)}), 
\end{eqnarray} 
here
$P_\ell(x)$ is the Legendre polynomial of order $\ell$. As
$\mp(R,\cos(\theta))$ is a symmetric function of $\cos(\theta)$, all
coefficients $\mp^\ell(R)$ for odd values of $\ell$ vanish. 

$m_0(R,\cos(\theta))$ is plotted in \Fig{fig:m0Rcosth}(a) for
$\Fr=0.1$ and four different values of $\St$, for $R=0.5$.
$m^0_0(R)$, $m^2_0(R)$, and $m^4_0(R)$ are plotted in
\Fig{fig:m0Rcosth}(b) for $\St=3.02$ and $\Fr=0.1$.  We find that
$m^\ell_0(R) = C_\ell R^{\zeta_\ell}$. Exponents $\zeta_\ell$ are
independent of $\ell$ and are equal to $\dtwo$ whereas the coefficients
$C_\ell$ is zero for odd $\ell$ and decrease with $\ell$ for even
$\ell$. 

To measure the degree of anisotropy we plot $m^2_0/m^0_0$ as a function of
$R$ for different values of $\St$ in \Fig{fig:m0Rcosth}(c). We observe
that this ratio does not depend on $R$ for the range of $R$ shown
here. This is because both $m^2_0$ and $m^0_0$ scales with the same
power as a function of $R$. We also find that anisotropy increases
with increasing $\St$.

In \Fig{fig:m1Rcosth}, we repeat this analysis for
$m_1(R,\cos(\theta))$. \Fig{fig:m1Rcosth} (b) shows plots for
$m^0_1$ and $m^2_1$ as a function of $R$ for $\St=3.03$ and $\Fr =
0.1$. Both of these coefficients show a power-law behaviour with
roughly the same exponent. This is more clear from the plots of
$m^2_1/m^0_1$ as a function of $R$ shown in \Fig{fig:m1Rcosth} (c). It
can be seen that the curves for higher values of $\St$ are not
constant as a function of $R$, but dependence is very weak.      

\subsection{Mean collision velocity} \label{subsec:colvel}
\begin{figure*} \begin{center}
\includegraphics[width=0.32\linewidth]{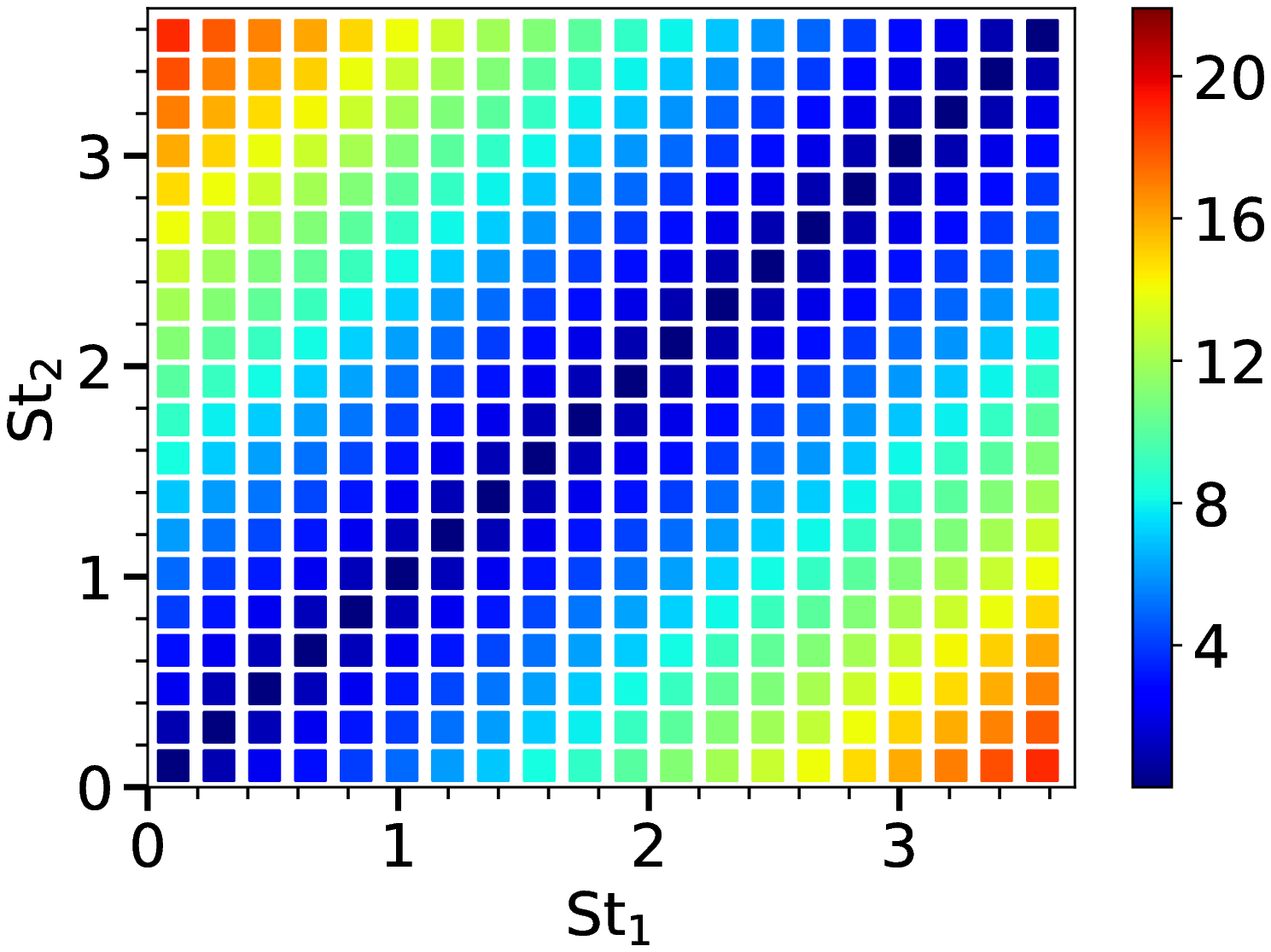}
\includegraphics[width=0.32\linewidth]{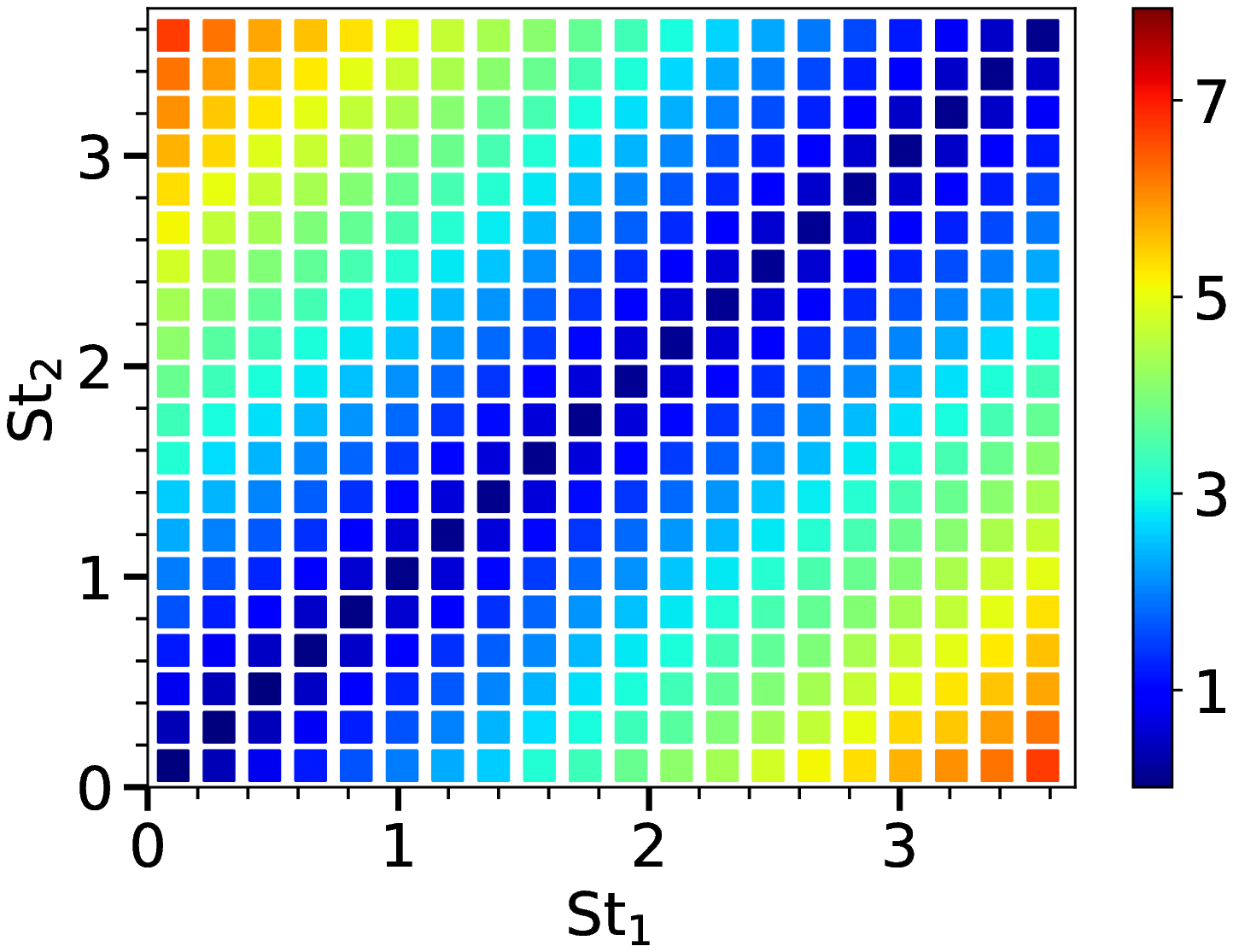}
\includegraphics[width=0.32\linewidth]{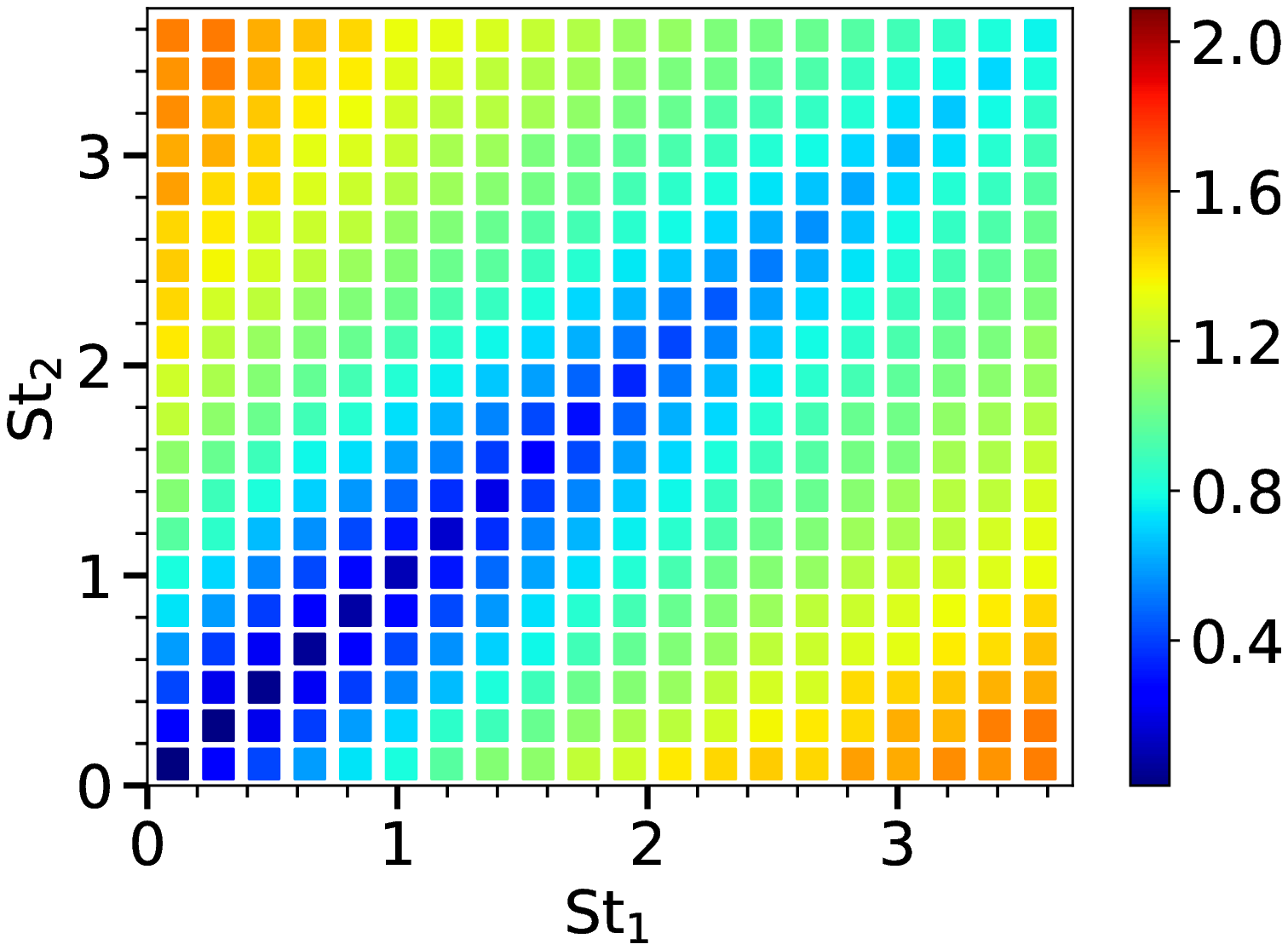} 
\caption{(color online) Mean $\VRn$  plotted as
a function of $\Sto$ and $\Stt$, for $\Fr = 0.1$ (left), $\Fr = 0.3$
(middle), and $\Fr = \infty$ (right).} 
\label{fig:mvSt1St2}
\end{center} \end{figure*}

The mean rate of collision depends on the relative velocity of particles when the separation 
between their centre of masses is equal to the sum of their radii. We define this as
collision velocity $\VRn$ (see \Eq{eq:mVn} and text before it). 
In \Fig{fig:mvSt1St2} we plot the mean of $\VRn$  
as a function of $\Sto$ and $\Stt$, for $\Fr=0.3,0.1$, and $\infty$.
We observe that the qualitative nature of these plot changes due to
gravity.  Along the diagonal $\Sto=\Stt$, values for non zero gravity are
smaller compared to  no gravity case. Away from diagonal for different
sized particles mean of $\VRn$ becomes a function of
$|\Sto-\Stt|$ for the settling particles. We also notice that values
for settling particles away from the diagonal are much higher compared
to particles with $\Fr=\infty$.

\begin{figure} \begin{center}
\includegraphics[width=1.0\linewidth]{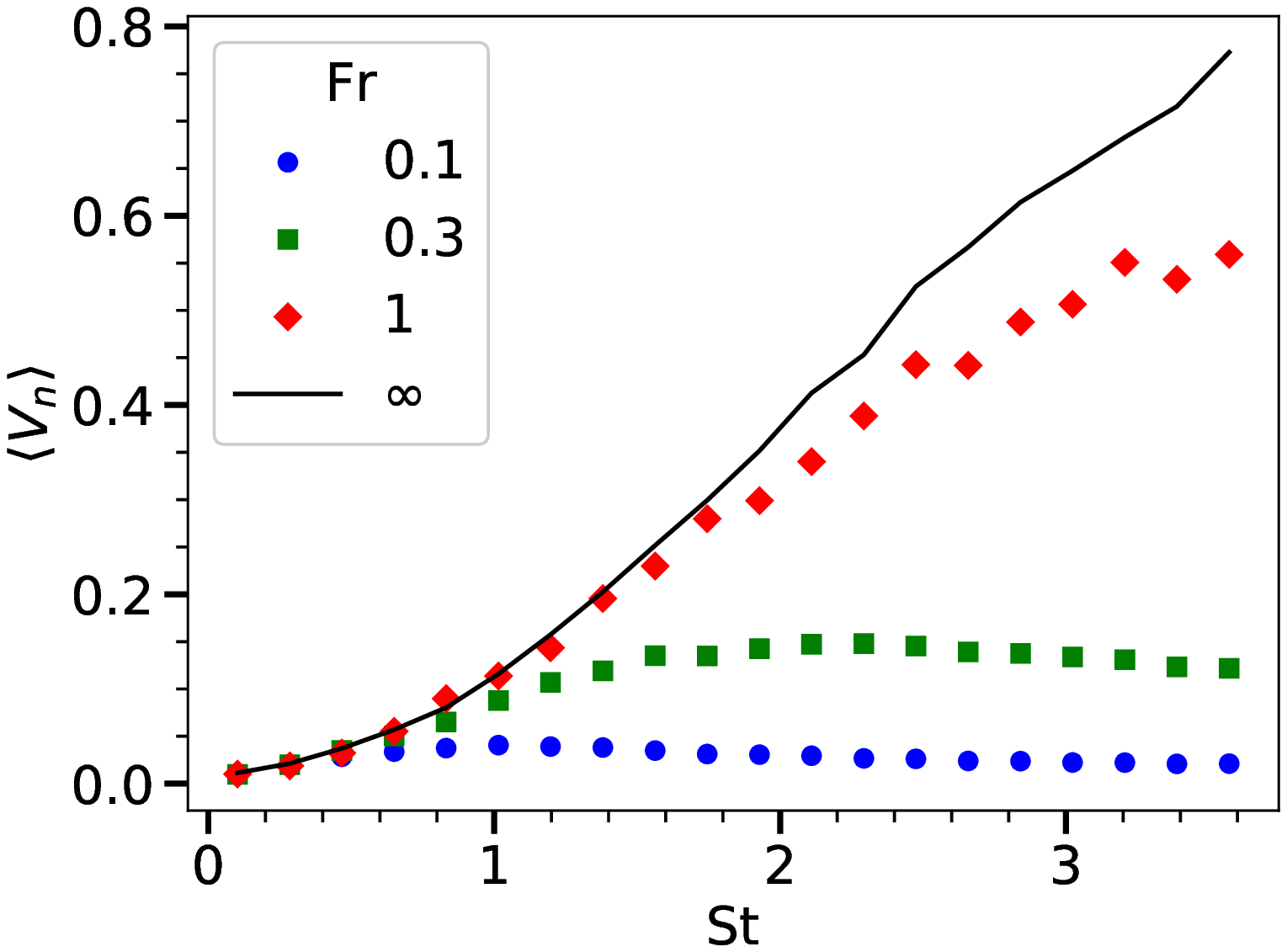}
\put(-30,150){(a)}\\
\includegraphics[width=1.0\linewidth]{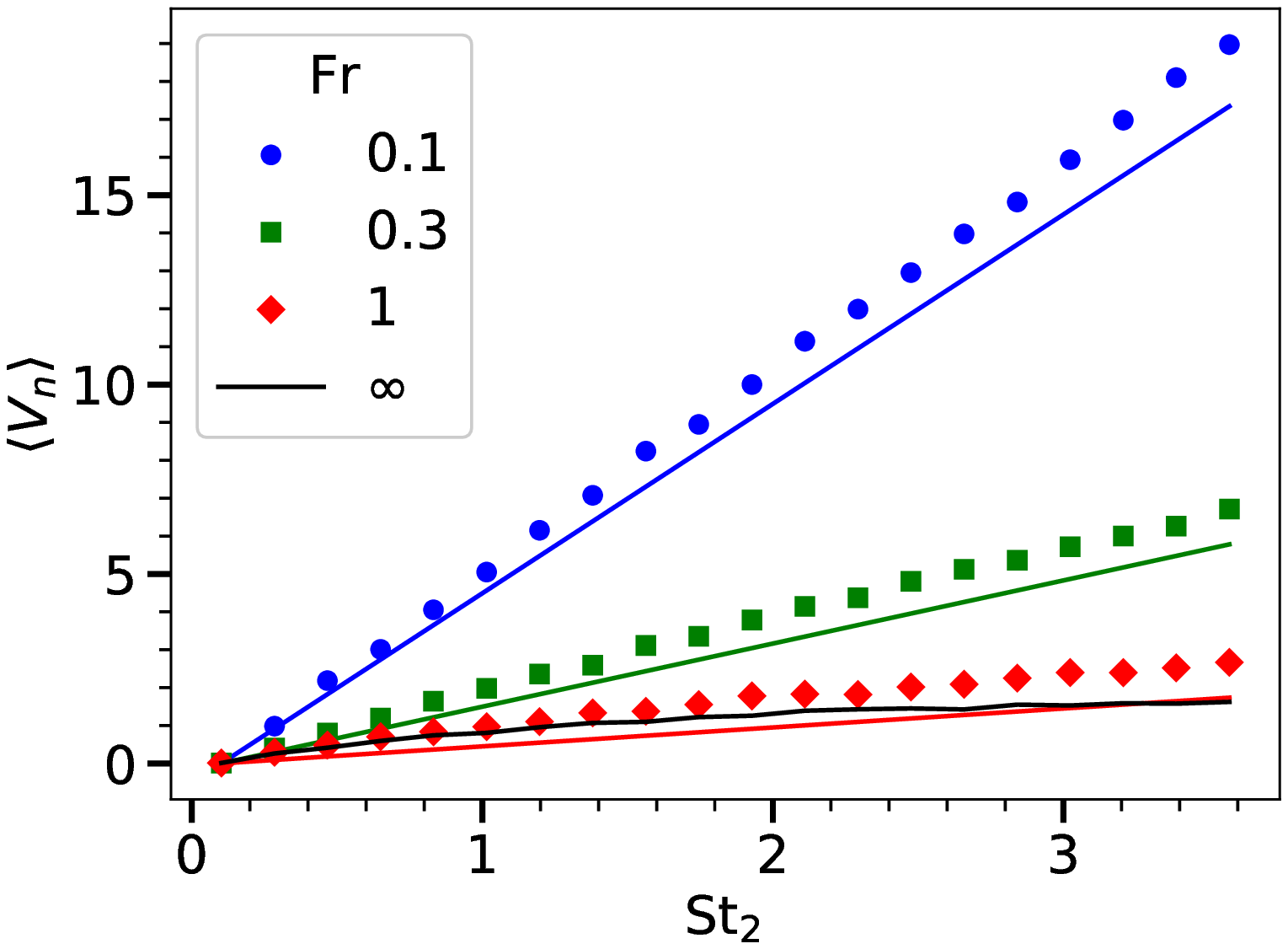}
\put(-30,150){(b)} \caption{(color online) Mean of $\VRn$ at $R = a_1
+ a_2$ for $\Sto = \Stt$ (top) and for $\Sto = 0.1$ as a function of
$\Stt$ (bottom), for different values of $\Fr$.  }
\label{fig:mVdiag+edge} \end{center} \end{figure}

To analyze it more quantitatively, we plot $\bra{\VRn}$ for
$\Sto=\Stt=\St$ in \Fig{fig:mVdiag+edge} (a), here $\bra{}$ denotes the mean. 
We observe that as $\Fr$
decrease $\bra{\VRn}$ decrease. In \Fig{fig:mVdiag+edge} (b) we plot
$\bra{\VRn}$ as a function of $\Stt$ for $\Sto=0.1$. In this case, we
find that $\bra{\VRn}$ increases significantly as $\Fr$ is decreased.
Solid lines in this plot show the $\bra{\VRn}$ for particles settling 
under gravity in a fluid at rest. 

\section{Conclusions} \label{sec:con} 
We have studied the effect of
gravity on the relative velocities of particles suspended in 
turbulent flow by using direct numerical simulations.  We explored a
range of $\St$ and $\Fr$ that is relevant for small droplets in
clouds. Our results show that joint PDFs of $R$ and $\VRa$ has a form
that is qualitatively similar to the one obtained without gravity. It
is shown that for separations smaller than the Kolmogorov scale $\eta$
($R < 1$ in nondimensional units), there exist a scale $\zast$ such
that joint PDF is independent of $\VRa$ for $\VRa < \zast R$, and is
independent of $R$ for $\VRa > \zast R$. We also show that PDFs of
$\VRa$ for a fixed value of $R$ scales as $\VRa^{\Dtwo-d-1}$ for some
range of $\VRa$. The phase space correlation dimension $\Dtwo$ that
measures the clustering of the particles in position-velocity
phase-space is modified by the presence of gravity. For $\St>1$,
$\Dtwo$ has a smaller value for $\Fr=0.3$ and $0.1$ compared to
$\Dtwo$ for $\Fr=\infty$.  This implies that the position- and
phase-space clustering is increased by the gravity for $\St>1$. 

We compute the moments of these joint PDFs of $R$ and $\VRa$ as a
function of $R$ and angle $\theta$ between the separation vector $\RR$
and the direction of gravity. We showed that the moments depend on the
$\cos{(\theta)}$ for a constant value of $R$. This indicates
anisotropy in the system due to the presence of gravity. To characterize
this anisotropy, we use a spherical harmonic decomposition and
compute coefficients of order $\ell=0,2,$ and $4$.  We found that all
these coefficient have a power-law dependence on $R$. Exponents of the
power-law is found to be the same for all values of $\ell$ that can be
obtained for $0$-th and $1$-th order moments. We define the degree of
anisotropy as the ratio of $\ell=2$ and $\ell=0$ coefficients. As
both the coefficients scale with the same exponent, degree of anisotropy
is independent of $R$, for $R<1$. A similar analysis of anisotropy is
done in Ref.~\cite{ire+bra+col16} for $R>1$. This study shows that for
larger $R$, the degree of anisotropy depends on $R$ and goes to $0$ for
very large values of $R$. 
 
We define the velocity of collision $\VRn$ as the radial component of the
relative velocity of two particles separated by the sum of their
radii, when two particles are approaching each other. We calculated
the mean of $\VRn$ as a function of $\Sto$ and $\Stt$ and found that
it changes qualitatively from the zero-gravity case. Due to the presence
of gravity mean of $\VRn$ becomes a function of $\mid \Sto-\Stt \mid$
alone.  Furthermore, for particles with equal $\St$, mean $\VRn$
decreases as $\Fr$ decreases, i.e., gravity increases. This decrease
is more pronounced at higher $\St$ than at lower ones.  For particle
having different values of $\St$ the qualitative behaviour is opposite,
the mean collision velocity increase as $\Fr$ decreases. For $\Sto
\neq \Stt$ collision velocity is determined by the difference in the
settling velocities of two particles. 

Joint PDFs of $R$ and $\VRa$ in the presence of gravity are studied
for the first time in this paper.
Refs.~\cite{pari+aya+ros+wan+gra15,ire+bra+col16} have studied the
PDFs of $\VR$ for small values of $R$ but power-law nature of the PDFs
is not shown. Real space correlation dimension $\dtwo$ for settling
particles is studied in Ref.~\cite{bec+hom+ray14}. This study does not
take into account the anisotropy due to the presence of gravity. Our study
shows that the scaling exponents $\dtwo$ for anisotropic contributions
are the same, but the amplitudes are different. Our results also show that
gravity can have a significant effect on the collision velocities of
particles having $\St$ between $0.5$ and $3$ and hence should be taken
in to account in the studies relevant for cloud droplets in this range
of $\St$.       

\section{Acknowledgments}
We thank Dhrubaditya Mitra, K. Gustafsson, B. Mehlig, and J. Bec for useful discussions.  
This work is supported by the grant Bottlenecks
for particle growth in turbulent aerosols from
the Knut and Alice Wallenberg Foundation (Dnr. KAW
2014.0048), by  Vetenskapsradet [grants 2013-3992 and 2017-03865], and Formas [grant number 2014-585].
Computational resources were provided by the
Swedish National Infrastructure for Computing (SNIC)
at PDC. 

\bibliographystyle{apsrev4-1} 
\bibliography{ref,turb_ref,stlng_part}

\end{document}